\documentstyle[12pt]{dubna98}
\input psfig
\begin{document}
\begin{center}
{\bf PION--PION SCATTERING EXPERIMENTS AT LOW ENERGY\\}
\vspace*{1cm}
DINKO PO\v{C}ANI\'C\\
{\it Department of Physics, University of Virginia, Charlottesville VA
22901 USA} 
\end{center}

\vspace*{.5cm} 
\begin{abstract} 

{\small This work reviews the available experimental information on the
$\pi$-$\pi$ scattering lengths, especially the recent near-threshold $\pi
N\to\pi\pi N$ data from several laboratories and the related application of
the Chew--Low--Goebel (CLG) technique well below 1\,GeV/c momentum.  At
this time uncertainties stemming from non-pion-exchange backgrounds in
near-threshold CLG studies appear to preclude a determination of the
$\pi$-$\pi$ scattering lengths with the desired accuracy of 10\% or
better.}

\end{abstract}

\vspace*{1cm}

\section{Motivation \label{sec:motiv}}

Pion-pion scattering at threshold is uniquely sensitive to the explicit
chiral symmetry breaking (ChSB) part of the strong interaction.  For this
reason it has been the subject of much theoretical and experimental study
for over thirty years.  While QCD has removed the early controversies and
established the Weinberg picture\cite{wein:1} of ChSB as valid at the tree
level, pion-pion scattering lengths, $a(\pi\pi)$, remain of interest in
terms of improving the precision of several basic parameters of the
effective chiral lagrangian.

Currently the most accurate predictions of $a(\pi\pi)$ come from chiral
perturbation (ChPT) calculations including terms up to two
loops.\cite{bij:col:1} However, a complementary approach, the generalized
chiral perturbation theory\cite{ste:etal:1} (GChPT) dispenses with the
standard assumption of a strong scalar quark condensate $\langle
0|\bar{q}q| 0\rangle$, allowing it to vary widely, with the conclusion that
available experimental evidence favors a fairly weak
condensate.\cite{kne:mou} The only observables capable of resolving this
discrepancy are the $\pi$-$\pi$ s-wave scattering lengths, with a required
precision of $\sim$\,10\,\% or better.

The result generally regarded as most reliable among the available
evaluations of $a_l^I(\pi\pi)$ was obtained in 1979 in a comprehensive
phase shift analysis\cite{nag:etal} of peripheral $\pi N\to\pi\pi N$
reactions and $K_{\rm e4}$ decays: 
\begin{equation}
    a_0^0 = 0.26 \pm 0.05\,\mu^{-1} \qquad {\rm and} \qquad 
    a_0^2 = -0.028 \pm 0.012\,\mu^{-1} \enspace , \label{eq:nagels}
\end{equation}
where $\mu$ is the pion mass, clearly not precise enough to resolve the
above quark condensate controversy.  We therefore examine the more recent
experiments and attempts at extraction of new, more precise values of
$a(\pi\pi)$.

\section{Experiments on Threshold $\pi$-$\pi$ Scattering}

Since free pion targets cannot be fabricated, experimental evaluation of
$\pi\pi$ scattering observables is restricted to the study of a dipion
system in a final state of more complicated reactions.  While several
reactions have been proposed and/or studied, only $\pi N\to\pi\pi N$ and
$K_{e4}$ decays have so far proven useful in studying threshold $\pi\pi$
scattering, although there are ambitious plans to study $\pi^+\pi^-$ atoms
(pionium) in the near future.  
\medskip

\hbox to \textwidth{\it \hfill $K_{\rm e4}$ Decays \hfill} 
\smallskip

By most measures, the $K^+\to\pi^+\pi^-e^+\nu$ decay (called $K_{\rm e4}$)
provides the most suitable tool for the study of threshold $\pi\pi$
interactions.  The interacting pions are real and on the mass shell, the
only hadrons in the final state.  The dipion invariant mass distribution
peaks close to the $\pi\pi$ threshold, and only two states,
$l_{\pi\pi}=I_{\pi\pi}=0$ and $l_{\pi\pi}=I_{\pi\pi}=1$, contribute
appreciably.  These factors, plus the well understood $V-A$ weak lagrangian
giving rise to the decay, favor the $K_{\rm e4}$ process among all others
in terms of theoretical uncertainties.  Measurements are, however, impeded
by the low branching ratio of the decay, $3.9 \times 10^{-5}$.

$K_{\rm e4}$ decay data provide information on the $\pi$-$\pi$ phase
difference $\delta^0_0 - \delta^1_1$ near threshold.  The most recent
published $K_{\rm e4}$ experimental result was obtained by a Geneva--Saclay
collaboration in the mid-1970's.\cite{ros:etal} Taken alone these data
provide a $\sim 35$\,\% constraint on $a_0^0$.  Only after being combined
with $\pi\pi$ phase shifts extracted from peripheral $\pi N\to\pi\pi N$
reactions (see below) is it possible to reduce the uncertainties to the
level of about 20\,\%, as quoted in Eq.\ (\ref{eq:nagels}).

We note that $K_{\rm e4}$ decays provide no information on the $I=2$
$\pi\pi$ phase shifts, meaning that information from other reactions is
required.
\medskip

\hbox to \textwidth{\it \hfill Peripheral $\pi N\to\pi\pi N$ Reactions at
High Momenta \hfill} 
\smallskip

Goebel as well as Chew and Low showed in 1958/59 that particle production
in peripheral collisions can be used to extract information on the
scattering of two of the particles in the final state.\cite{che:low} This
approach is, of course, useful primarily for the scattering of unstable
particles and has been used to great advantage in the study of the $\pi\pi$
system.  The method relies on an accurate extrapolation of the double
differential cross section to the pion pole, $t=\mu^2$ ($t$ is the square
of the 4-momentum transfer to the nucleon), in order to isolate the one
pion exchange (OPE) pole term contribution.  Since the exchanged pion is
off-shell in the physical region ($t<0$), this method requires measurements
under conditions that maximize the OPE contribution and minimize all
background contributions---typically peripheral pion production at values of
$t$ as close to zero as possible, which is practical at incident
momenta above $\sim$3\,GeV/c.  Since the CLG method relies on
extrapolation in a two-dimensional space, it requires kinematically
complete data of high quality, both in terms of measurement statistics and
resolution---main limiting factors in all analyses to date.

The data base for peripheral CLG analyses has not changed essentially since
the early 1970's, and is dominated by two experiments, performed by the
Berkeley\cite{pro:etal} and CERN-Munich\cite{gra:etal} groups.  A
comprehensive analysis of this data base, with addition of the
Geneva--Saclay $K_{\rm e4}$ data, was performed by Nagels et
al.\cite{nag:etal}, with results given in Eq.\ (\ref{eq:nagels}).

A 1982 analysis by the Kurchatov Institute group was based on a set of some
35,000 $\pi N\to\pi\pi N$ events recorded in bubble
chambers.\cite{ale:etal} This analysis was recently updated by including
available data on the $\pi N\to\pi\pi\Delta$ reaction, as well as the
published $K_{\rm e4}$ data.\cite{pat:etal} The resulting s-wave $\pi\pi$
scattering lengths were bounded by
\begin{equation}
   0.205\,\mu^{-1} < a_0^0 < 0.270\,\mu^{-1} \quad {\rm and} \quad
   -0.048\,\mu^{-1} < a_0^2 < -0.016\,\mu^{-1} \enspace . \label{eq:patarak}
\end{equation}
Although the limits on $a_0^0$ carry slightly smaller uncertainties than
the $a_0^0$ value of Nagels et al.\ given in Eq.\ (\ref{eq:nagels}), the
result of Patarakin et al.\ still cannot rule out any of the two competing
pictures of chiral symmetry breaking (strong vs.\ weak scalar quark
condensate).  The central value, though, is lower than in
(\ref{eq:nagels}), more in line with the conventional, strong condensate  
picture that leads to the standard ChPT two-loop prediction of $a_0^0
\simeq 0.21\,\mu^{-1}$.

New high energy ($E_\pi > 3$\,GeV) peripheral $\pi N\to\pi\pi N$
measurements have not been planned for some time, so that much attention
has been devoted to the study of the $\pi N\to\pi\pi N$ reaction at lower
energies, $p_\pi \leq 500$\,MeV, as discussed below.
\medskip
 
\hbox to \textwidth{\it \hfill Inclusive $\pi N\to\pi\pi N$ Reactions Near
Threshold \hfill} 
\smallskip
 
Weinberg showed early on\cite{wein:1} that the OPE graph dominates the $\pi
N\to\pi\pi N$ reaction at threshold, inspiring vigorous theoretical and
experimental study of the $\pi\pi$ and $\pi N\to\pi\pi N$ threshold
amplitudes.  Results of near-threshold $\pi\pi N$ studies published before
1995 are reviewed in detail in Ref.\,\cite{poc:1}).  That impressive data
base has been augmented by the addition of new, more precise $\pi^\pm
p\to\pi^\pm\pi^+n$ cross sections very near threshold from
TRIUMF.\cite{lan:etal} The new measurements have confirmed the same group's
earlier published data\cite{sev:etal} on the $\pi^+p\to\pi^+\pi^+n$
reaction, thus definitively invalidating older data taken by the OMICRON
collaboration at CERN.\cite{kern:etal}

Notwithstanding the abundance and high accuracy of recent near-threshold
inclusive pion production data, their interpretation in terms of $\pi\pi$
amplitudes has been plagued by theoretical uncertainties.  This shortcoming
was addressed in 1995 using the heavy baryon chiral perturbation
theory\cite{ber:etal} (HBChPT), yielding:
\begin{equation}
    a_0^0 \simeq  0.21 \pm 0.07 \,\mu^{-1} \qquad {\rm and} \qquad
    a_0^2 = -0.031 \pm 0.007 \,\mu^{-1} \enspace .  \label{eq:hbchpt}
\end{equation}
The above result for $a_0^0$ was subsequently refined by Olsson who used
the so-called universal curve, a model-independent relation between $a_0^0$
and $a_0^2$ due to the forward dispersion relation or, equivalently, to the
Roy equations.\cite{olsson}  Olsson found 
\begin{equation}
    a_0^0 = 0.235 \pm 0.03 \,\mu^{-1} \enspace .      \label{eq:olsson}
\end{equation}
Any analysis based on HBChPT cannot, however, be expected to result in
$\pi\pi$ scattering lengths significantly different from the standard ChPT
prediction because the latter is built into the lagrangian used.
\medskip

\hbox to \textwidth{\it \hfill Chew--Low--Goebel Analysis of Low Energy
$\pi N\to\pi\pi N$ Data \hfill} 
\smallskip

Given the theoretical uncertainties in the interpretation of inclusive $\pi
N\to\pi\pi N$ data near threshold, it was suggested some time ago to
apply the Chew--Low method to low energy $\pi N\to\pi\pi N$
data.\cite{poc:etal:1}  Recently several exclusive $\pi N\to\pi\pi N$ data
sets suitable for such treatment have become available.  These are, in
the order in which they were measured: \\
\indent (a) $\pi^-p\to\pi^0\pi^0n$ data from BNL\cite{low:etal}, \\
\indent (b) $\pi^+p\to\pi^+\pi^0p$ data from LAMPF\cite{poc:etal:2}, and \\  
\indent (c) $\pi^-p\to\pi^-\pi^+n$ data from TRIUMF\cite{kerm:etal}. \\
We next briefly review the current results of these experiments.

A Virginia--Stanford--LAMPF team studied the $\pi^+p\to\pi^+\pi^0p$
reaction at LAMPF at five energies from 190 to 260\,MeV.\cite{poc:etal:2}
The LAMPF $\pi^0$ spectrometer and an array of plastic scintillation
telescopes were used for $\pi^+$ and $p$ detection.  Three classes of
exclusive events were recorded simultaneously and independently:
$\pi^+\pi^0$ and $\pi^0p$ double coincidences, and $\pi^+\pi^0p$ triple
coincidences.  The $\pi^+p\to\pi^+\pi^0p$ reaction is sensitive only to the
$I=2$ s-wave $\pi\pi$ scattering length.

The main source of difficulty in this analysis was the relatively broad
missing mass resolution: $\sigma_p \simeq 11$\, MeV and $\sigma_\pi \simeq
17$\, MeV.  This energy resolution considerably smears the cross section
data bins in a Chew--Low plot of $m_{\pi\pi}$ against $t$.  Consequently,
in order to obtain a physically interpretable array of double differential
cross section bins, a complicated deconvolution procedure had to be
implemented first.\cite{bru} Limited counting statistics presented an
additional difficulty in the analysis, as it increased the uncertainties in
both the deconvolution procedure and in the final CLG extrapolation.

\begin{figure}[htb]
\vbox{\hspace*{0.2\textwidth}
\psfig{figure=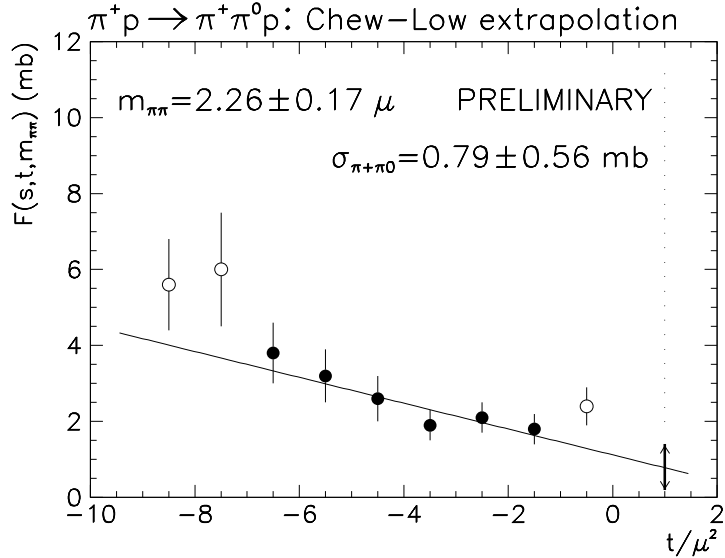,width=0.6\textwidth} }
\caption{\small Chew--Low extrapolation to the pion pole from $\pi^+p
\to \pi^+\pi^0p$ exclusive cross sections at 260 MeV (preliminary).  Full
circles: data points included in the fit.  Open circles: data points
excluded from the fit. The ex\-tra\-po\-la\-ted value of the $\pi\pi$ total
cross section at $m_{\pi\pi} = 2.26 \pm 0.18\ \mu$ is indicated.}
\label{fig:chew-low:e1179}
\end{figure}

Preliminary results of this analysis for one bin of $m_{\pi\pi} = 2.26 \pm
0.18\,\mu$ are shown in Fig.\,\ref{fig:chew-low:e1179}.  Open circles in
the figure indicate data points excluded from the Chew--Low extrapolation
procedure due to large values of $|t|>6\,\mu^2$, where OPE is weak, and the
smallest $|t|$ point which has a large normalization uncertainty due to the
cross section deconvolution procedure.  The resulting $\pi\pi$ cross
section is $0.79 \pm 0.56$\,mb, which translates to a phase shift of
$\delta_0^2 = -8.3^\circ \pm 3.0^\circ$.  This does not provide a strong
new constraint when compared with existing information.

\begin{figure}[htb]
\vbox{\hspace*{0.265\textwidth}
\psfig{figure=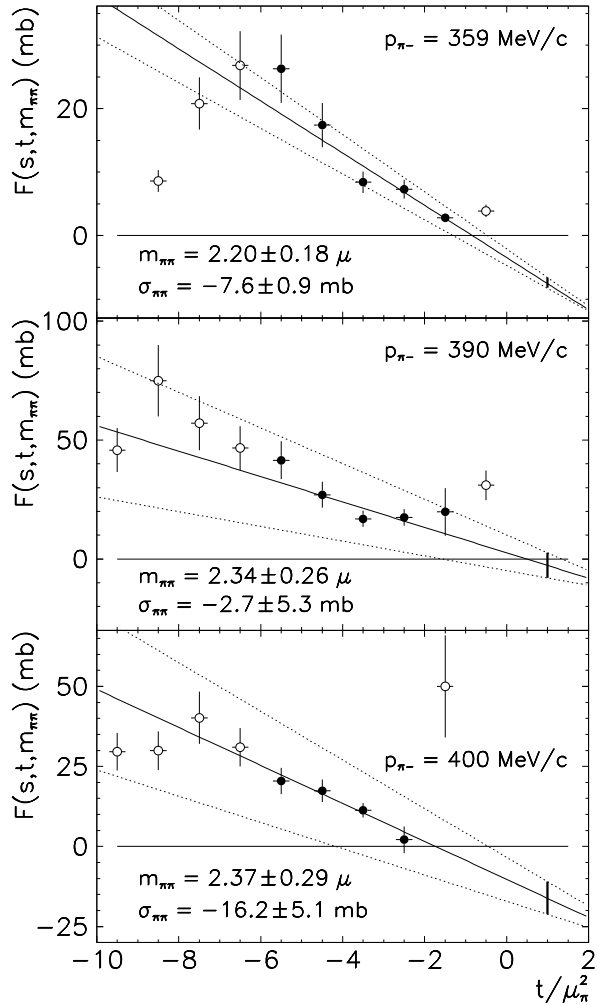,width=0.47\textwidth} }
\caption{\small Chew--Low extrapolation to the pion pole $t=+\mu^2$ based
on $\pi^-p \to \pi^0\pi^0n$ exclusive cross sections measured at three beam
momenta (preliminary).  Full circles: data points included in the fit.
Open circles: data points excluded from the fit. The unphysical negative
ex\-tra\-po\-la\-ted values of the $\pi\pi$ total cross section are
indicated.} 
\label{fig:chew-low:e857}
\end{figure}

In comparison, the BNL $\pi^-p\to\pi^0\pi^0n$ data, while having much
higher event statistics, are characterized by an even broader energy
resolution and poorer coverage of the low $|t|$ region critical for the
Chew--Low extrapolation.  This limitation and/or strong influence of
non-OPE backgrounds led to nonphysical results (negative extrapolated cross
sections), as shown in Fig.\,\ref{fig:chew-low:e857}.

The most significant development in this field in recent years has been the
construction and operation of the Canadian High Acceptance Orbit
Spectrometer (CHAOS), a sophisticated new detector at TRIUMF.\cite{chaos}
This impressive device, composed of a number of concentric cylindrical wire
chamber tracking detectors and total energy counters mounted between the
poles of a large bending magnet, provides nearly 360$^\circ$ of angular
coverage for in-plane events, with excellent acceptance for multi-particle
events.

\begin{figure}[htb]
\vbox{%\hspace*{0.05\textwidth}
\psfig{figure=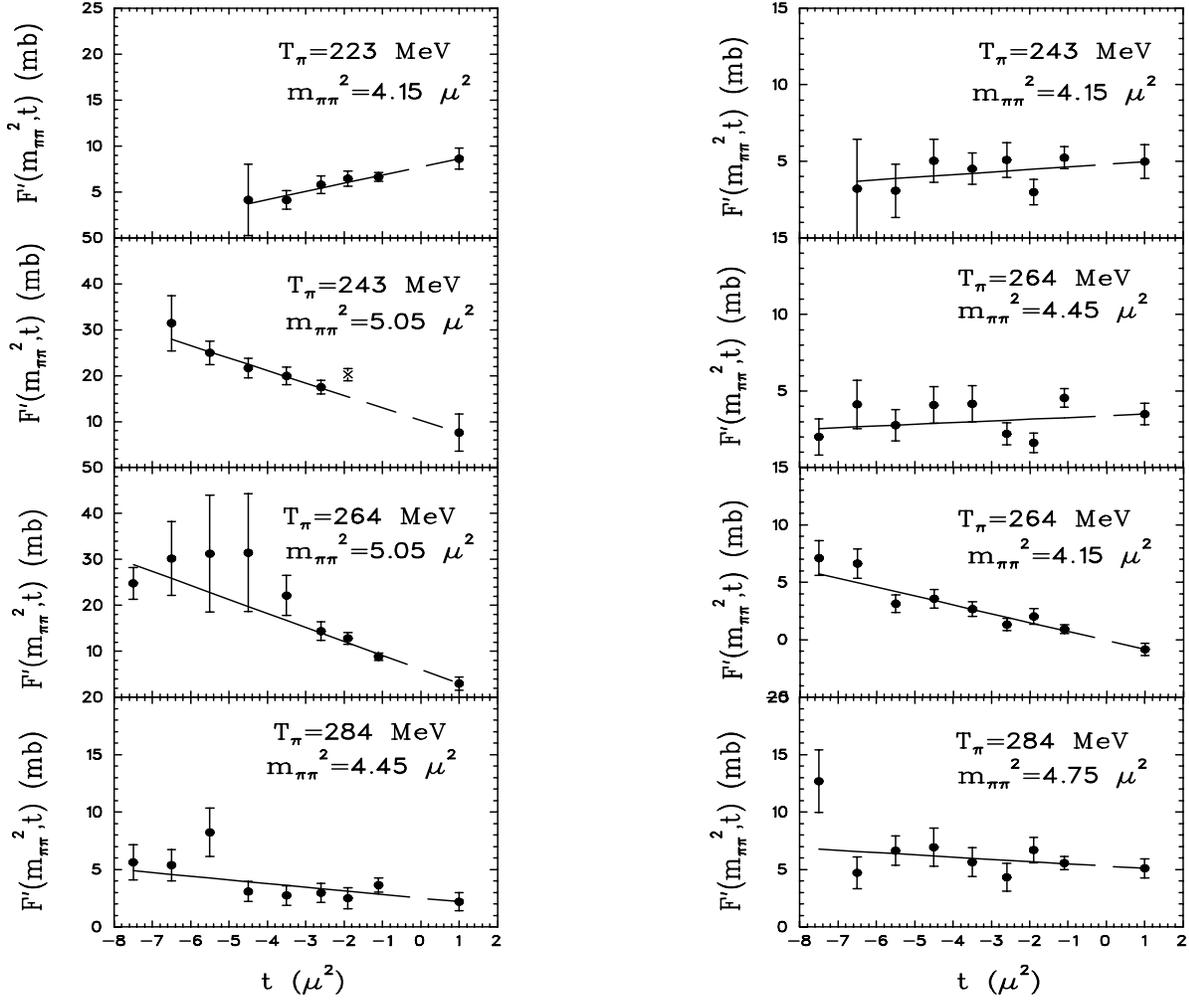,width=1.0\textwidth} }
\caption{\small Chew--Low extrapolation fits produced by the CHAOS
group from measured $\pi^-p\to\pi^+\pi^-n$ data.\protect\cite{kerm:etal}
The points at $t=+\mu$ are deduced from extrapolation and yield the
$\pi\pi$ cross section.  Solid circles: data points used in the linear fit;
crosses: data points not used in the fit.
\label{fig:chew-low:chaos} }
\end{figure}

The CHAOS $\pi^-p\to\pi^+\pi^-n$ data set covers four incident beam
energies between 223 and 284\,MeV.  Unlike the LAMPF and BNL measurements,
these data have an excellent energy resolution of $\sigma\simeq 4.8$\,MeV,
resulting in good linear Chew--Low extrapolations, as shown in
Fig.\,\ref{fig:chew-low:chaos}.

From the CLG fits the authors extracted
$\pi\pi$ cross sections at six $\pi\pi$ energies in the range $m_{\pi\pi}^2
= 4.15$--5.65\,$\mu^2$ with uncertainties ranging from about 16\,\% at the
lowest energy to 63\,\% at the highest.  These $\pi\pi$ cross section data
were then added to the data base of Ref.\,\cite{pat:etal}), and a Roy
equation constrained phase shift analysis was performed following the same
procedure as in Ref.\,\cite{pat:etal}), allowing $a_0^0$ to vary freely.
Minimizing the $\chi^2$ of the fit, the authors obtained
\begin{equation}
     a_0^0 = 0.206 \pm 0.013 \, \mu^{-1} \enspace , \label{eq:chaos_res}
\end{equation}
which would strongly confirm the validity of the standard ChPT and the
strong scalar quark condensate implied therein, at the same time ruling out
the possibility of the weak scalar quark condensate proposed by the Orsay
group.\cite{ste:etal:1}

However, Bolokhov et al.\ of the Sankt Petersburg State University have
recently performed a detailed study of the reliability of the Chew--Low
method at low energies using sets of synthetic $\pi N\to\pi\pi N$ ``data''
between 300 and 500 MeV/c.\cite{bol:etal}  In this work the authors
constructed data sets with: (a) the OPE contribution only, (b) OPE + other
allowed mechanisms, (c) all mechanisms without the OPE.  Both linear and
quadratic Chew--Low extrapolation were used.  The authors found 25--35\,\%
deviations in the reconstructed OPE strength in case (a), 100--300\,\%
deviations under (b), and large ``OPE amplitude'' without any pion pole in
the synthetic data under (c).  This led the authors to conclude that the
Chew--Low method appears to give completely unreliable results.  However,
given the complex nature of the issue, it would be premature to write off
using the method at low energies altogether.  Clearly, a critical
examination of the problem is imperative.  In the meantime, before the
matter is finally resolved, we cannot accept the CHAOS result in
Eq.\ (\ref{eq:chaos_res}) as definitive.

\section{Summary of Current Results and Future Prospects
\label{sec:compare}} 

\begin{figure}[htb] 
\vbox{\hspace*{0.15\textwidth}
\psfig{figure=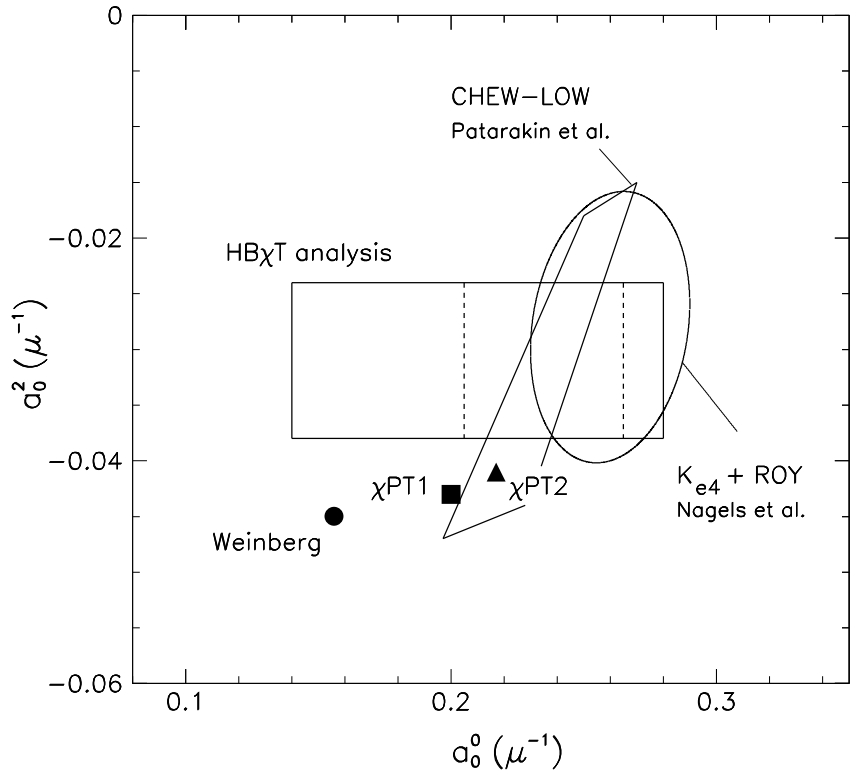,width=0.7\textwidth} } 
\caption{\small Summary of $\pi\pi$ scattering length predictions: Weinberg's
tree-level result\protect\cite{wein:1} (full circle), ChPT one-loop
calculation\protect\cite{gas:leu:2} (full square), ChPT two-loop
calculation\protect\cite{bij:col:1} (full triangle), and analyses of
experimental data: Nagels et al.\protect\cite{nag:etal} (oval contour),
Patarakin et al.\protect\cite{pat:etal} (oblique quadrangular contour),
HBChPT analysis of Bernard et al.\protect\cite{ber:etal} (solid
rectangle), and Olsson's dispersion-relation constraint of the HBChPT
result\protect\cite{olsson} (dashed lines).  
\label{fig:univ_plot} } 
\end{figure}

Theoretical predictions and experimental results on the $\pi\pi$ scattering
lengths published to date are plotted in Fig.\,\ref{fig:univ_plot} in the
$a_0^2$ against $a_0^0$ plane.  
It is clear that the current analyses of the available $K_{\rm e4}$ and $\pi
N\to\pi\pi N$ data (excluding the not yet fully established low energy
application of the Chew--Low method) are not sufficiently accurate to
distinguish between the two scenarios of chiral symmetry breaking, i.e.,
between the standard picture and the one with a weak $\langle
0|\bar{q}q|0\rangle$. 

At the same time the available analyses seem to favor slightly higher
values of both $a_0^0$ and $a_0^2$ than the values predicted by standard
ChPT (strong $\langle 0|\bar{q}q|0\rangle$).

The threshold $\pi$-$\pi$ scattering experimental data base will improve
significantly in the near future as several new experiments bear fruit.
These are: (a) the forthcoming $K_{\rm e4}$ data from BNL E865 (experiment
completed, analysis in progress) and the KLOE detector at DA$\Phi$NE
(experiment to start soon), as well as (b) the planned measurement of the 
lifetime of the $\pi^+\pi^-$ atom (the DIRAC project at CERN).  If all goes
as planned, these experiments combined will provide $\sim$\,5\,\% limits on
the scattering lengths.

As noted above, further theoretical work is required to make use of the
existing $\pi N\to\pi\pi N$ data, in particular to clarify the
applicability of the Chew--Low--Goebel method at low energies.
Additionally, better understanding of the electromagnetic corrections will
be necessary in order to take full advantage of the forthcoming $K_{\rm
e4}$ and pionium data.  Thus, the next few years will be interesting on
both the experimental and theoretical fronts.
\smallskip

The author wishes to thank A. A. Bolokhov, E. Frle\v{z}, O. O. Patarakin,
M. E. Sevior and G. R. Smith for substantive discussions and for graciously
providing access to results of their ongoing work.  This work has been
supported by a grant from the U.S. National Science Foundation.

% 
% ---- Bibliography ---- 
% 

\end{document}